\documentclass[reprint,aps,prl]{revtex4-1}    
\usepackage{graphicx}
\usepackage{natbib}
\usepackage{amsmath,amssymb}
\usepackage{pxfonts}
\usepackage{lmodern}
\usepackage{times}
\usepackage{bm}
\usepackage{epstopdf}
\usepackage{ulem}
\usepackage[colorlinks,linkcolor=blue,urlcolor=blue,citecolor=blue]{hyperref}
\hypersetup{
    colorlinks=true,
    linkcolor=blue,
    filecolor=magenta,
    urlcolor=cyan,
    citecolor=cyan
}
\begin{document}

\title{Synchronization through frequency shuffling}
\author{Manaoj Aravind$^1$}
\thanks{These authors contributed equally to this work}
\email[\\Corresponding author email: ]{manaojaravind@iitb.ac.in}
\email[\\Email: ]{sarkar@thphys.uni-heidelberg.de}
\author{Vaibhav Pachaulee$^1$}
\thanks{These authors contributed equally to this work}
\email[\\Corresponding author email: ]{manaojaravind@iitb.ac.in}
\email[\\Email: ]{sarkar@thphys.uni-heidelberg.de}
\author{Mrinal Sarkar$^2$}
\thanks{These authors contributed equally to this work}
\email[\\Corresponding author email: ]{manaojaravind@iitb.ac.in}
\email[\\Email: ]{sarkar@thphys.uni-heidelberg.de}
\author{Ishant Tiwari$^1$}
\altaffiliation[Present address: ]{\textit{School of Chemical and Biomolecular engineering, Georgia institute of Technology, Atlanta, Georgia 30332-0100, U.S.A.}}
\author{Shamik Gupta$^3$} 
\author{P. Parmananda$^1$}
\affiliation{$^1$Department of Physics, Indian Institute of Technology Bombay, Powai, Mumbai $400$ $076$, India
\\
$^2$Institute for Theoretical Physics,
University of Heidelberg,
Philosophenweg 19,
D-69120 Heidelberg, Germany\\
$^3$Department of Theoretical Physics, Tata Institute of Fundamental Research, Homi Bhabha Road, Mumbai $400$ $005$, India}
\date{\today}

\begin{abstract}
   A wide variety of engineered and natural systems are modelled as networks of coupled nonlinear oscillators. In nature, the intrinsic frequencies of these oscillators are not constant in time. Here, we probe the effect of such a temporal heterogeneity on coupled oscillator networks, through the lens of the Kuramoto model. To do this, we shuffle repeatedly the intrinsic frequencies among the oscillators at either random or regular time intervals. What emerges is the remarkable effect that frequent shuffling induces earlier onset (i.e., at a lower coupling) of synchrony among the oscillator phases. Our study provides a novel strategy to induce and control synchrony under resource constraints. We demonstrate our results analytically and in experiments with a network of Wien Bridge oscillators with internal frequencies being shuffled in time.  
\end{abstract}
\maketitle

Synchronization is crucial for proper functionality, survival, and adaptation at various length and time scales across disciplines, from biology to physics to real-world systems~\cite{strogatz2004sync}. Attaining synchronization often comes with unavoidable energy costs under limited resources. For instance, in Kai system underlying cyanobacterial circadian clock, energy dissipation drives the coupling between the oscillators, and synchronization is achieved beyond a critical dissipation~\cite{zhang2020energy}. Another instance is multi-agent network systems, e.g., sensor networks, distributed computation, multiple robot systems, where the control energy is limited, leading to tradeoffs between synchronization-regulation performance and energy budget~\cite{xi2018dynamic}. In systems modeled as interacting oscillators of distributed intrinsic frequencies, one constraint is the limited coupling budget~\cite{nishikawa2006maximum,zhang2021designing}, which might be insufficient for synchrony. Designing an optimal protocol to achieve synchronization with a given energy/coupling budget is of great practical relevance. Synchrony at low couplings has mostly been achieved via developing networks whose topology changes in time~\cite{zhang2021designing}. This Letter achieves the goal by introducing a novel protocol of \textit{shuffling} the oscillator frequencies, which we show to be inducing synchrony \textit{even in a static network in otherwise-unfavorable conditions.}

A compelling motivation for adopting a protocol such as ours stems from time variability of internal system-parameters. Examples include variability in neuronal firing activity in the brain, facilitating development of epileptic seizures~\cite{stefanescu2012computational}, time variability in cardiac and respiratory frequencies during anesthesia~\cite{musizza2007interactions}, cardiovascular signals containing oscillatory components with time-varying frequencies, e.g., effects of aging on heart rate variability~\cite{shiogai2010nonlinear}. Understanding effects of temporal variation of intrinsic frequencies gives insights into such systems, often modeled as non-autonomous systems with time-varying forcing~\cite{suprunenko2013chronotaxic, lucas2021synchronisation, petkoski2012kuramoto}. We offer an alternative way of incorporating \textit{temporal heterogeneity} in  internal parameters, by introducing shuffling of intrinsic frequencies.

\begin{figure}[ht]
    \centering
\includegraphics[width=0.85\linewidth]{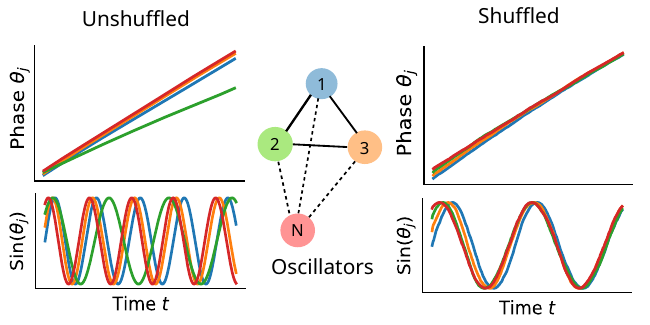}
    \caption{Onset of synchrony (synchronized temporal evolution of phases $\theta_j$ and $\sin(\theta_j)$'s) in the Kuramoto model (middle figure) on introducing frequency shuffling (right panels) even at coupling constant values at which the unshuffled system (left panels) does not support such a state.}
    \label{fig:scheme}
\end{figure}

\begin{figure*}[ht]
    \includegraphics[width=0.95\linewidth]{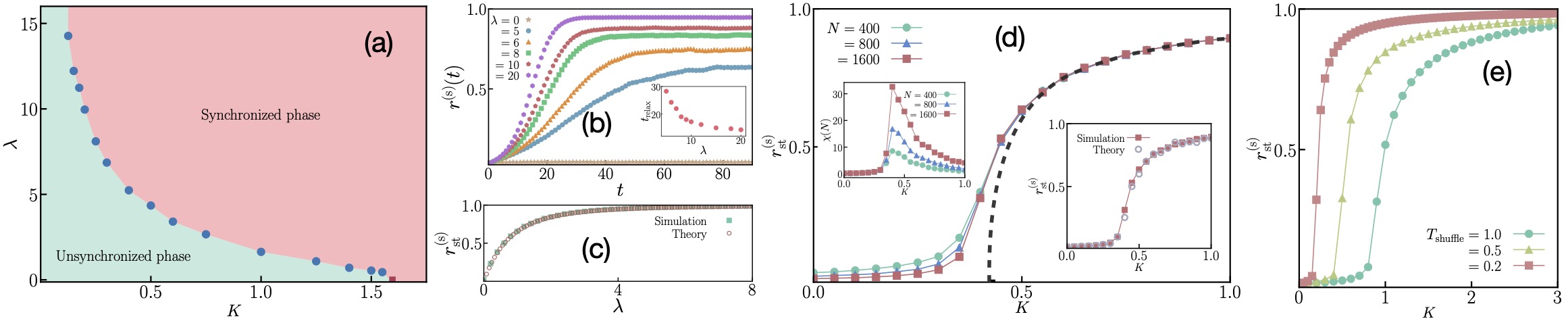}
     \caption{ \textbf{(a)} Phase diagram in the ($K,\lambda$)-plane for the Kuramoto model subject to shuffling at random times with a constant rate $\lambda$. The red square is the result for the bare Kuramoto dynamics. The boundary between the two depicted phases marks the critical coupling $K_c(\lambda)$. Alternatively, if we fix a value of $K$, then the critical $\lambda$ to observe synchronization is the value of $\lambda$ that lies on the phase boundary  corresponding to the given $K$. \textbf{(b)} Time evolution of the order parameter $r^{(\mathrm{s})}$, starting from a random initial condition, for no shuffling ($\lambda=0$) and shuffling with rates $\lambda = 5, 6, 8, 10$ and $20$. The system size is $N=800$ and coupling strength is $K=0.5$. \textbf{Inset}: $t_{\rm{relax}}$ (time for $r^{(\mathrm{s})} (t)$ to attain half of its stationary-state value) as a function of $\lambda$, for $K=0.5$ and $N=800$. \textbf{(c)}  Exact result (see text) and its numerical verification for the case $K=0$ and simultaneous shuffling and phase resetting at a constant rate $\lambda$. The system size is $N=1600$. \textbf{(d)} Stationary order parameter $r_{\rm {st}}^{(\mathrm{s})}$ versus $K$ for shuffling rate $\lambda = 5.0$. Three system-sizes: $N=400,~ 800$, and $1600$ are considered (the limiting behavior as $N \to \infty$ shown by the dashed line). \textbf{Inset (Left)}: Stationary-state fluctuations $\chi(N) \equiv N \left[ \langle (r_\mathrm{st}^{(\mathrm{s})})^2 \rangle - \langle r_\mathrm{st}^{(\mathrm{s})}\rangle^2 \right]$ versus $K$ for the same parameter values and system sizes, bearing a signature of a phase transition in the limit $N \to \infty$. \textbf{Inset (Right)}: Agreement between theory and simulations for $r_{\rm {st}}^{(\mathrm{s})}$, for $N = 1600$ and $\lambda=5.0$. \textbf{(e)} $r_\mathrm{st}^{(\mathrm{s})}$ vs. $K$ from simulations of the Kuramoto model subject to shuffling at regular intervals of duration $T_\mathrm{shuffle}=0.2, ~0.5$ and $1.0$. The system size is $N=1600$. In all cases, the frequency distribution is a Gaussian with zero mean and unit variance.}
    \label{fig:theory_sim}
\end{figure*}

We develop a framework in the ambit of the paradigmatic model of synchronization~\cite{strogatz2004sync}: the Kuramoto model of coupled phase-only oscillators with distributed frequencies~\cite{pikovsky2001synchronisation,gupta2018statistical,kuramoto1984chemical,strogatz2000kuramoto,acebron2005kuramoto, gupta2014kuramoto,verma2013potential, verma2015kuramoto, bera2017coexisting, temirbayev2013autonomous, english2015experimental,parmananda2001resonance, chigwada2006resonance, zhang2021random, sugitani2021synchronizing, nicolaou2020coherent, nair2021using}. Here, we ask: What happens when the Kuramoto model undergoing its bare evolution is interspersed with instantaneous \textit{frequency-shuffling} at random times with a constant shuffling rate $\lambda>0$? Shuffling involves a redistribution of frequencies among the oscillators. Our main message is analytical and experimental demonstration that shuffling leads to the remarkable effect of an earlier onset of synchronization and reduction in the value of critical $K$ to observe synchronization: the shuffled system synchronizes even when the corresponding unshuffled system does not (see the phase diagram in the $(K,\lambda)$-plane in Fig.~\ref{fig:theory_sim}(a), where $\lambda \to 0$ refers to the unshuffled system)! In the bare Kuramoto model, the coupling needs to be finely tuned beyond a threshold to observe synchrony, whereas shuffling when done frequently-enough leads to synchrony at arbitrary coupling. These features conform to the objective of attaining synchronization under coupling budget. Not just that, the time $t_{\rm{relax}}$ to achieve synchronization starting from an unsynchronized state has a dramatic consequence of shuffling: $t_{\rm{relax}}$ for a given $K$ decreases with increasing $\lambda$ (Fig.~\ref{fig:theory_sim}(b)). Thus, synchronization under shuffling gets easier in every practical sense: one requires smaller coupling and shorter time to achieve synchrony. Shuffling at random times is not restrictive, as we show similar results on shuffling at fixed time-intervals. We experimentally demonstrate synchronization under shuffling using Wien Bridge oscillators.

Synchronization in our system is induced by the noise introduced by shuffling in the otherwise-deterministic dynamics. Noise-induced synchronization among limit-cycle oscillators has been extensively studied~\cite{teramae2004robustness, goldobin2005synchronization, nakao2007noise, yoshimura2007synchronization, nagai2010noise, pinto2016thermodynamics, kawamura2016optimization}. The basic framework involves Langevin dynamics for the oscillator-variables that includes a common Gaussian white noise $\xi(t)$ (or colored noise) acting on all oscillators: (i) ${\rm d} \bold{X}_j/{\rm d}t =  \bold{F}(\bold{X}_j) +  \boldsymbol{\xi}(t)$, or, (ii) for phase-only oscillators, ${\rm d} \theta_j/{\rm d}t=  \omega_j +  \epsilon f(\theta_j) {\xi}(t)$ and ${\rm d} \theta_j/{\rm d}t=  \omega_j +  (K/N) \sum_{k=1}^{N} \sin(\theta_k - \theta_j) + \sin(\theta_j)\xi(t)$ for Kuramoto oscillators. This results in an effective time-varying intrinsic frequency given by $\omega_j +  \epsilon f(\theta_j) {\xi}(t)$ and $\omega_j +  \sin(\theta_j)\xi(t)$, respectively,  yielding incremental variation of effective frequencies in time: in a small time $\mathrm{d}t$, only a small change $\propto \sqrt{\mathrm{d}t}$ (in addition to  a ``noise-induced drift" term $\propto \mathrm{d}t$, for (ii)) takes place. This case is restrictive and very different from our case in which the frequencies undergo large changes in $\mathrm{d}t$ only at shuffling instants, making the dynamics piecewise-deterministic. Obviously, the emerging behavior will be different in the two cases: The theme of piecewise-deterministic frequency-changes inducing synchronization remains unexplored to date, highlighting the novelty of our contribution.

Figure~\ref{fig:scheme} shows our results schematically. For weak coupling, the unshuffled system has phases $\theta_j$ of individual oscillators with different intrinsic frequencies growing independently in time and $\sin(\theta_j)$'s varying periodically in time with different frequencies. Even at such a low $K$, when the unshuffled system is unsynchronized, shuffling remarkably leads to synchronization in oscillator phases and in the $\sin(\theta_j)$'s.   

The Kuramoto model comprises $N$ globally-coupled oscillators with distributed intrinsic frequencies. The phase $\theta_j \in [0,2\pi)$ of the $j$-th oscillator, $j=1,2,\ldots,N$, evolves as
\begin{equation}
\frac{\mathrm{d}\theta_j}{\mathrm{d}t}=\omega_j+\frac{K}{N}\sum_{k=1}^N \sin(\theta_k-\theta_j).
\label{eq:eom}
\end{equation}
Here, $K \ge 0$ is the coupling, while $\omega_j$ denotes the intrinsic frequency. The frequencies $\{\omega_j\}$ are independent and identically-distributed random variables following a given distribution $G(\omega);~\int_{-\infty}^\infty \mathrm{d}\omega~G(\omega)=1$. We take $G(\omega)$ to have zero mean (effect of any finite mean vanishes by going to a co-rotating frame) and finite variance $\sigma^2$. The effect of the latter is made explicit by rewriting Eq.~\eqref{eq:eom} as $\mathrm{d}\theta_j/\mathrm{d}t=\sigma\omega_j+(K/N)\sum_{k=1}^N \sin(\theta_k-\theta_j)$,
and by taking $\{\omega_j\}$'s to follow a distribution $g(\omega)$ with zero mean and unit variance. Under rescaling: $K \to K' \equiv K/\sigma$, $t \to t' \equiv t\sigma$ and $\lambda \to \lambda' \equiv \lambda/\sigma$, and omitting the primes, the rescaled equation has the same form as Eq.~\eqref{eq:eom}, with the $\{\omega_j\}$'s sampled from a distribution $g(\omega)$ with zero mean and unit variance. We refer to this set-up as the bare Kuramoto model. In this bare model, the set $\{\omega_j\}$ is constructed once at time $t=0$ and is fixed throughout the temporal evolution. In the model with shuffling, wherein the frequencies are repeatedly shuffled and redistributed among the oscillators after random time intervals, the definition of the shuffling rate $\lambda$ implies that the time interval $\tau$ between successive shuffling is distributed as an exponential $p(\tau)=\lambda e^{-\lambda \tau}$; the average time between two successive shuffling is $\langle \tau \rangle=1/\lambda$. The bare model is recovered in the limit $\lambda \to 0$, and which as $N \to \infty$ and at long times $t\to \infty$ (stationary state) exhibits a phase transition in the order parameter $R(t)=r(t)e^{\mathrm{i}\psi(t)}=(1/N)\sum_j e^{\mathrm{i}\theta_j(t)}$, between a low-$K$ unsynchronized phase ($r_\mathrm{st}\equiv r(t\to \infty)=0$) and a high-$K$ synchronized phase ($0 < r_\mathrm{st} \leq 1$) across the critical value $K_c=2/ \pi g(0)$~\cite{strogatz2000kuramoto}. As detailed below, a finite $\lambda$ leads to a rich stationary-state phase diagram in the $(K,\lambda)$-plane (e.g., Fig.~\ref{fig:theory_sim}(a) for Gaussian $g(\omega)$), with a synchronized phase in a region where the bare model does not support such a phase and with the critical coupling $K_c (\lambda)$ most notably decreasing with increasing $\lambda$. 

To analyse the model with shuffling in the limit $N \to \infty$, we define a conditional probability density $P^{(\mathrm{s})}_\omega(\theta,t|\theta',t')$ to find an oscillator with frequency $\omega$ that has phase $\theta$ at time $t$, conditioned on having found an oscillator with the same frequency and with phase $\theta'$ at an earlier instant $t'<t$. Here, the superscript ``s" stands for shuffling. The normalization reads as $\int_0^{2\pi}\mathrm{d}\theta~P_\omega^{(\mathrm{s})}(\theta,t|\theta',t')=1~\forall~\omega,\theta',t,t'<t$, while the order parameter reads $R^{(\mathrm{s})}(t)=r^{(\mathrm{s})}(t)e^{\mathrm{i}\psi^{(\mathrm{s})}(t)}=\int \mathrm{d}\theta \mathrm{d}\omega~g(\omega)P_\omega^{(\mathrm{s})}(\theta,t|\theta',t')e^{\mathrm{i}\theta}$. Note that shuffling implies that the oscillator(s) with frequency $\omega$ and phase $\theta$ at time $t$ could be different from the one(s) with the same frequency but with phase $\theta'$ at time $t'<t$. This is unlike the case in absence of shuffling, when the oscillators have the same frequency throughout the evolution. Note that the number of oscillators with a given frequency is conserved in time both in presence and absence of shuffling. In the latter case, the corresponding probability density $P_\omega(\theta,t|\theta',t')$ evolves as~\cite{gupta2018statistical} 
\begin{align}
    \frac{\partial P_\omega(\theta,t|\theta',t')}{\partial t}=-\frac{\partial }{\partial \theta}\left[\left(\omega+Kr(t)\sin(\psi(t)-\theta)\right)P_\omega\right],
    \label{eq:continuity-equation}
\end{align}
with $R(t)=r(t)e^{\mathrm{i}\psi(t)}=\int \mathrm{d}\theta \mathrm{d}\omega~g(\omega)P_\omega(\theta,t|\theta',t')e^{\mathrm{i}\theta}$.
Knowing $P_\omega(\theta,t|\theta',t')$ yields $P_\omega^{(\mathrm{s})}(\theta,t|\theta_0,0)$, with $P_\omega^{(\mathrm{s})}(\theta,0|\theta_0,0)=\delta(\theta-\theta_0)$ describing the (given) initial condition $\theta=\theta_0$ for all oscillators, and using renewal theory~\cite{cox1962renewal}, as
\begin{align}
&P^{(\mathrm{s})}_\omega=e^{-\lambda t}P_\omega(\theta,t|\theta_0,0)+\lambda \int_0^t \mathrm{d}\tau~e^{-\lambda \tau}P_\omega(\theta,t|\theta(t'),t'),
\label{eq:renewal-equation}
\end{align}
with $t'=t-\tau$. Indeed, for the given initial condition, the probability to observe an oscillator with phase $\theta$ and frequency $\omega$ at time $t$ requires the dynamics to either (i)  not have undergone a single shuffling since the initial time instant $t=0$, or, (ii) have the last shuffling during the time interval $[t-\tau-\mathrm{d}\tau,t-\tau];~\tau \in [0,t]$, and thereafter free evolution up to time $t$ starting with the phase value $\theta(t-\tau)$ attained at time instant $t-\tau$ under the dynamical evolution. The first and second terms on the right hand side (RHS) of Eq.~\eqref{eq:renewal-equation} correspond to the cases (i) and (ii), respectively. The function $P_\omega(\theta,t|\theta(t-\tau),t-\tau)$ needs to be constructed by letting the $N$ oscillators undergo the Kuramoto dynamics interspersed with shuffling at random times, from $t=0$ to $t=t$, and noting the fraction of oscillators that at time $t$ have frequency $\omega$ and phase $\theta$, and which have undergone the last shuffling at time instant $t-\tau$ (to frequency $\omega$) when their phase value was $\theta(t-\tau)$.  One obtains from Eq.~\eqref{eq:renewal-equation} the order parameter as
\begin{align}
R^{(\mathrm{s})}(t)|_{\theta_0}&= e^{-\lambda t}R(t)|_{\theta_0} + \lambda \int_0^t \mathrm{d}\tau~e^{-\lambda \tau}R(t)|_{\theta(t-\tau)}.
\label{eq:r_renewal-equation}
\end{align}
Here, $R(t)|_{\theta(t-\tau)}$ is the value of $R(t)$ under dynamical evolution according to the bare Kuramoto model and with $\theta(t-\tau)$ as the initial condition. On the other hand, $R^{(\mathrm{s})}(t)|_{\theta_0}$ is the value of $R^{(\mathrm{s})}(t)$ under dynamical evolution according to the Kuramoto model with shuffling and with $\theta_0$ as the initial condition.
As $t \to \infty$, Eq.~\eqref{eq:r_renewal-equation} yields the stationary value:
\begin{align}
R^{(\mathrm{s})}_\mathrm{st}=r_\mathrm{st}^{(\mathrm{s})}e^{\mathrm{i}\psi^{(\mathrm{s})}_\mathrm{st}}= \lim_{t \to \infty}\lambda \int_0^t \mathrm{d}\tau~e^{-\lambda \tau}R(t)|_{\theta(t-\tau)}.  
\label{eq:renewal-shuffling}
\end{align}
The RHS of the second equality does not depend on $\theta_0$.
Equation~\eqref{eq:renewal-shuffling} will prove  crucial in obtaining our main results. Note that Eqs.~\eqref{eq:renewal-equation},~\eqref{eq:r_renewal-equation},~\eqref{eq:renewal-shuffling} are very general and apply to \textit{any} $g(\omega)$.

Proceeding further requires to know $P_\omega(\theta,t|\theta',t')$, which being $2\pi$-periodic in $\theta$ admits the Fourier expansion $P_\omega(\theta,t|\theta',t')=\sum_{n=-\infty}^\infty \widetilde{P}_n^{(\omega)}(t|\theta',t')e^{\mathrm{i}n\theta}$;~$P_\omega$ being real, $[\widetilde{P}_n^{(\omega)}]^\star=\widetilde{P}_{-n}^{(\omega)}$, with star denoting complex conjugation. Equation~\eqref{eq:continuity-equation} yields 
\begin{align}
\frac{\partial \widetilde{P}_n^{(\omega)}}{\partial t}=-\mathrm{i}n\omega \widetilde{P}_n^{(\omega)}-\frac{r(t)Kn}{2}(e^{\mathrm{i}\psi(t)}\widetilde{P}^{(\omega)}_{n+1}-e^{-\mathrm{i}\psi(t)}\widetilde{P}_{n-1}^{(\omega)}), 
\label{eq:Fourier_coeff_evolution}
\end{align}
with $r(t)e^{\mathrm{i}\psi(t)}=\int \mathrm{d}\omega~g(\omega)\widetilde{P}_{-1}^{(\omega)}$. Equation~\eqref{eq:Fourier_coeff_evolution} is a non-linear integro-differential equation, whose solution is difficult to obtain in general. Before proceeding, it proves worthwhile to analyse the special case of non-interacting oscillators.

For non-interacting oscillators ($K=0$), the solution of Eq.~\eqref{eq:Fourier_coeff_evolution}, with the obvious condition $P_\omega(\theta,t'|\theta',t')=\delta(\theta-\theta')$, is $\widetilde{P}_n^{(\omega)}(t|\theta',t')=e^{-\mathrm{i}(n\omega(t-t')+n\theta')}/(2\pi)$. We get $P_\omega(\theta,t|\theta',t')=\sum_{n=-\infty}^\infty e^{\mathrm{i}n(\theta-\theta'-\omega(t-t')}/(2\pi)$, giving $R(t)|_{\theta(t-\tau)}=\int \mathrm{d}\omega~g(\omega)~e^{\mathrm{i}(\omega \tau+\theta(t-\tau))}$, and Eq.~\eqref{eq:renewal-shuffling} yielding
\begin{equation}
r_\mathrm{st}^{(\mathrm{s})}e^{\mathrm{i}\psi^{(s)}_\mathrm{st}}=\lim_{t\to \infty}\lambda \int_0^t \mathrm{d}\tau \int \mathrm{d}\omega g(\omega)e^{-\lambda \tau+\mathrm{i}(\omega \tau+\theta(t-\tau))},
    \label{eq:Rst-equation}
\end{equation}
which still requires to evolve the oscillators under Kuramoto dynamics with $K=0$ and interspersed with shuffling to determine $\theta(t-\tau)$. Things simplify for the special case of phase resetting, wherein at random time intervals, together with frequency shuffling, the oscillator phases are reset to a common value $\Theta$. Here, all memory of previous time evolution is washed out following every shuffling. This corresponds to setting $\theta(t-\tau)=\Theta$, yielding from Eq.~\eqref{eq:Rst-equation}: $\psi^{(\mathrm{s})}_\mathrm{st}=\Theta$,
$r^{(\mathrm{s})}_\mathrm{st}=\lambda \int \mathrm{d}\omega~g(\omega)/(\lambda - \mathrm{i}\omega)$,
applicable to any $g(\omega)$.
With even $g(\omega)$: $g(\omega)=g(-\omega)$, one gets $r^{(\mathrm{s})}_\mathrm{st}=2\lambda^2 \int_0^\infty \mathrm{d}\omega~g(\omega)/(\lambda^2+\omega^2)$. For example, for Gaussian $g(\omega)$ with zero mean and unit variance, we get $r^{(\mathrm{s})}_\mathrm{st}=\lambda \sqrt{\pi/2}~e^{\lambda^2/2}\mathrm{Erfc}(\lambda/\sqrt{2})$, with $\mathrm{Erfc}(x)$ the complementary error function. We have thus an exact result on the stationary-state order parameter for non-interacting oscillators subject to simultaneous shuffling and phase resetting. An excellent agreement between this exact result and simulations, see Fig.~\ref{fig:theory_sim}(c), is a vindication of our theoretical approach.

We now analyse shuffling among interacting oscillators ($K\ne 0$). In the absence of an analytical solution for $\theta$ as a function of time, Eq.~\eqref{eq:renewal-shuffling} may be evaluated semi-analytically by using for a large-enough $N$ the data for $\theta(t-\tau)$ obtained from simulations of the Kuramoto model with shuffling, and the results may be compared against simulation results for validation; the proposed analysis is very general and applies to any $g(\omega)$. For the representative case of a Gaussian $g(\omega)$ with zero mean and unit variance~\cite{note},
Fig.~\ref{fig:theory_sim}(d)[right inset] shows a very good match between theory and simulation results for $N=1600$. Such a good match with theory applies not just to Fig.~\ref{fig:theory_sim}(d)[right inset], but to all our numerical results. The main figure in Fig.~\ref{fig:theory_sim}(d) shows the behavior of $r^{(\mathrm{s})}_\mathrm{st}$ for three different $N$, with the black dashed line corresponding to the behavior as $N \to \infty$. A remarkable feature implied by the figure is a phase transition as $N \to \infty$ from an unsynchronized ($r^{(\mathrm{s})}_\mathrm{st}=0$) to a synchronized ($r^{(\mathrm{s})}_\mathrm{st}\ne 0$) phase at a critical $K$ that depends on $\lambda$. 

In simulations for a given $\lambda$, use of finite $N$ rounds off the phase transition point $K_c$, as is well known from theory of phase transitions~\cite{Fisher1967,binder-book}. This requires finite-size scaling analysis to extract the ``true" phase transition point occurring as $N\to \infty$. To this end, one considers the quantity $\chi(N) \equiv N \left[ \langle (r_\mathrm{st}^{(\mathrm{s})})^2 \rangle - \langle r_\mathrm{st}^{(\mathrm{s})}\rangle^2 \right]$, measuring stationary-state fluctuations of the order parameter; here, angular brackets denote averaging over dynamical realizations. Concomitant with the existence of a phase transition at $K_c(\lambda)=K_c(\lambda,N\to \infty)$ as $N \to \infty$ occurs the divergence of the quantity $\chi(N\to \infty)$ at $K_c(\lambda)$. For finite $N$, instead, a peak in $\chi(N)$ at a ``pseudo"-critical point $K_c(\lambda,N)$ is observed, with the peak value increasing with increasing $N$, see Fig.~\ref{fig:theory_sim}(d)[left inset] for Gaussian $g(\omega)$. Consequently, the quantity $K_c(\lambda)$ is obtained by identifying $K_c(\lambda,N)$ as the point at which $\chi(N)$ peaks and then by plotting $K_c(\lambda,N)$ versus $N$ to extract the $N\to \infty$ behavior. The result for $K_c(\lambda)$ as a function of $\lambda$ is shown in the phase diagram in Fig.~\ref{fig:theory_sim}(a) for Gaussian $g(\omega)$. In the limit $\lambda\to 0$, when the system has the bare Kuramoto evolution, the critical $K$ is given by the Kuramoto model result $K_c=2/(\pi g(0))=2\sqrt{2/\pi}$. We see that $K_c(\lambda)$ decreases monotonically with increasing $\lambda$. Thus, introducing shuffling dramatically alters the phase diagram of the bare model: the shuffled system admits a synchronized phase even when the bare model does not support such a phase and synchronizing gets easier with shuffling, the main message of this work. The phase diagram in Fig.~\ref{fig:theory_sim}(a) suggests that as $\lambda \to \infty$, the critical $K$ approaches zero. Said differently, for non-interacting oscillators, a non-zero $r^{(\mathrm{s})}_\mathrm{st}$ is possible only in the trivial limit $\lambda \to \infty$ (continuous shuffling). Aside from this trivial case of non-interacting oscillators, which cannot be called synchronization~\cite{pikovsky2001synchronisation}, our main message nevertheless is what Fig.~\ref{fig:theory_sim}(a) depicts: For interacting oscillators ($K \neq 0$), shuffling dramatically reduces the critical $K$ to observe synchronization,  facilitating its occurrence. Until now, we have considered shuffling at a constant rate $\lambda$. Figure~\ref{fig:theory_sim}(e) for shuffling after regular time interval $T_\mathrm{shuffle}$ shows qualitatively similar results: shuffling facilitates synchronization.  
\begin{figure}
    \centering
    \includegraphics[width=\linewidth]{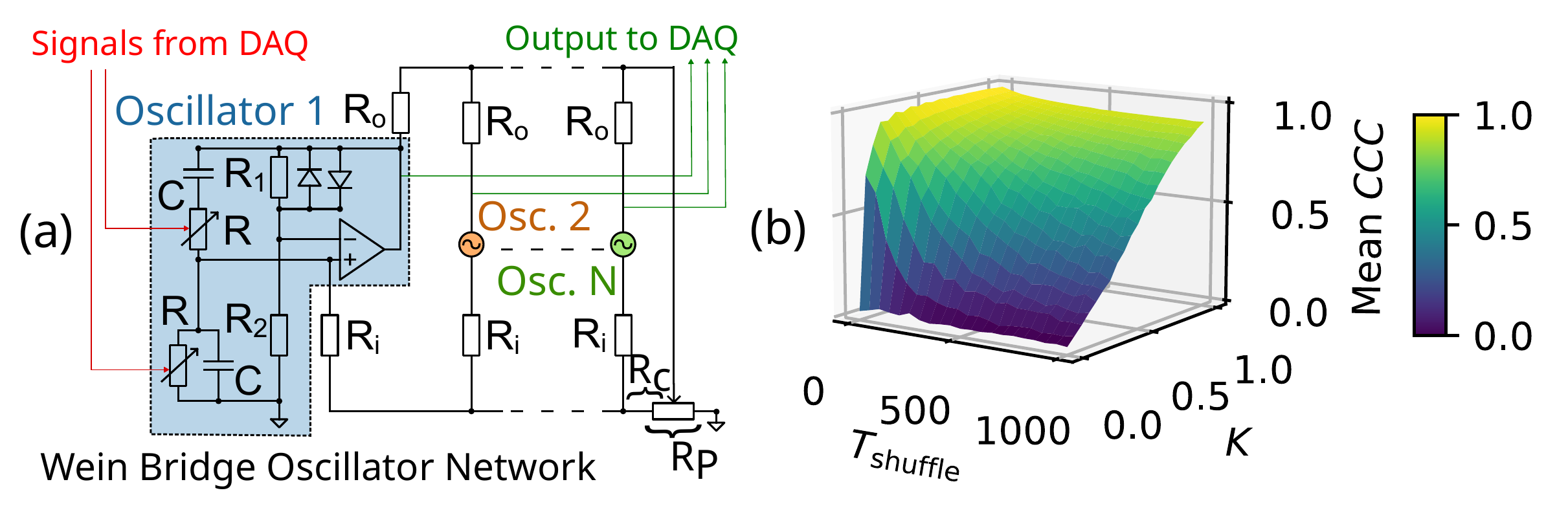}
    \caption{(a) Schematic  experimental set-up. Six coupled Wien Bridge (WB) oscillators (one of them shown within dashed lines) with frequencies varied in real time through voltage-dependent resistors $R$. The red lines depict incoming voltage signals to an oscillator and the green lines show the output voltage signals from each oscillator. The collected voltage data was analysed, and the results are shown in panel (b). Here, $C = 100$ nF, $R_1 = 3.3$ K$\Omega$, $R_2 = 1$ K$\Omega$, $R_i = 1$ K$\Omega$, $R_o = 30$ K$\Omega$. A potentiometer $R_P$ of $10$ K$\Omega$ controls the coupling $K$. (b) Mean cross-correlation coefficient (CCC) among the six WB oscillator outputs, as a function of $K$ and shuffling interval $T_\mathrm{shuffle}$. }
    \label{fig:circuit_heatmap}
\end{figure}

 To demonstrate our results experimentally, we construct a Wien Bridge (WB) oscillator-network studying synchronization~\cite{temirbayev2013autonomous,english2015experimental, english2016emergence}; see
 Fig.~\ref{fig:circuit_heatmap}(a). A globally-coupled network of $N=6$ WB oscillators was built. The coupling strength $K$ is controlled by a potentiometer. A dimensionless measure of $K$ is obtained knowing the resistance $R_c$ and the potentiometer resistance $R_P$: $K=(R_P-R_c)/R_P$. The intrinsic frequency of each WB oscillator depends on the capacitance $C$ and the resistance $R$. To vary this frequency, voltage-dependent resistors~\cite{melby2005dynamics} were used to vary $R$. To control these resistors, voltage signals sampled uniformly in $[1.5 - 2.5 V]$ were generated and transmitted through a high-speed data acquisition device (DAQ) - Measurement Computing USB 1616HS. This generated oscillations with frequency in $[250 - 350]$ Hz. The voltage outputs from individual oscillators were continually recorded using the same DAQ. The DAQ was set up to collect and relay voltages at a sampling rate $1.5 \times 10^5$ samples/second. The shuffling interval $T_\mathrm{shuffle}$ is measured in multiples of sampling time interval. The frequency of each oscillator stays constant under a constant voltage applied to $R$. After a time interval $T_\mathrm{shuffle}$, a new voltage is chosen uniformly in $[1.5-2.5]$, effecting sampling of frequencies from the same distribution. This routine when extended to large $N$ is equivalent to repeated shuffling of the frequencies after time $T_\mathrm{shuffle}$. After a long runtime, recorded voltage outputs of the six oscillators were analysed. We measure synchrony by means of the mean cross-correlation coefficient (CCC), i.e., the mean of all pairwise Pearson correlation coefficients. Figure~\ref{fig:circuit_heatmap}(b) shows a surface plot of mean $CCC$ versus $K$ and $T_\mathrm{shuffle}$. The transition to synchrony shifts to lower $K$ for more frequent shuffling. This experimental demonstration points to the robustness and generality of easier synchronization with frequency shuffling. Given small $N=6$, the setup demonstrates shuffling-induced synchronization not only for infinite, but also for finite (small) $N$, despite fluctuations of the average frequency at every frequency-sampling event.

In this work, effect of shuffling of frequencies was studied in coupled nonlinear oscillators: under shuffling, at regular or random time intervals, synchronization is achieved at arbitrary coupling, provided one shuffles frequently enough. Our work offers a new leash for real applications, especially, in inducing and controlling synchronization under resource constraints, besides providing a framework to study time-varying oscillator networks. 

\begin{acknowledgments}
M.S. acknowledges support from the Deutsche Forschungsgemeinschaft (DFG, German Research Foundation) under Germany’s Excellence Strategy EXC 2181/1-390900948 (the Heidelberg STRUCTURES Excellence Cluster). SG acknowledges support from  SERB-CRG Grant CRG/2020/000596 and ICTP, Trieste, Italy, for
support under its Regular Associateship scheme.
\end{acknowledgments}

\bibliographystyle{apsrev4-2.bst}

\begin{thebibliography}{0}%
\makeatletter
\providecommand \@ifxundefined [1]{%
 \@ifx{#1\undefined}
}%
\providecommand \@ifnum [1]{%
 \ifnum #1\expandafter \@firstoftwo
 \else \expandafter \@secondoftwo
 \fi
}%
\providecommand \@ifx [1]{%
 \ifx #1\expandafter \@firstoftwo
 \else \expandafter \@secondoftwo
 \fi
}%
\providecommand \natexlab [1]{#1}%
\providecommand \enquote  [1]{``#1''}%
\providecommand \bibnamefont  [1]{#1}%
\providecommand \bibfnamefont [1]{#1}%
\providecommand \citenamefont [1]{#1}%
\providecommand \href@noop [0]{\@secondoftwo}%
\providecommand \href [0]{\begingroup \@sanitize@url \@href}%
\providecommand \@href[1]{\@@startlink{#1}\@@href}%
\providecommand \@@href[1]{\endgroup#1\@@endlink}%
\providecommand \@sanitize@url [0]{\catcode `\\12\catcode `\$12\catcode
  `\&12\catcode `\#12\catcode `\^12\catcode `\_12\catcode `\%12\relax}%
\providecommand \@@startlink[1]{}%
\providecommand \@@endlink[0]{}%
\providecommand \url  [0]{\begingroup\@sanitize@url \@url }%
\providecommand \@url [1]{\endgroup\@href {#1}{\urlprefix }}%
\providecommand \urlprefix  [0]{URL }%
\providecommand \Eprint [0]{\href }%
\providecommand \doibase [0]{https://doi.org/}%
\providecommand \selectlanguage [0]{\@gobble}%
\providecommand \bibinfo  [0]{\@secondoftwo}%
\providecommand \bibfield  [0]{\@secondoftwo}%
\providecommand \translation [1]{[#1]}%
\providecommand \BibitemOpen [0]{}%
\providecommand \bibitemStop [0]{}%
\providecommand \bibitemNoStop [0]{.\EOS\space}%
\providecommand \EOS [0]{\spacefactor3000\relax}%
\providecommand \BibitemShut  [1]{\csname bibitem#1\endcsname}%
\let\auto@bib@innerbib\@empty
\end{thebibliography}%


\begin{thebibliography}{41}%
\makeatletter
\providecommand \@ifxundefined [1]{%
 \@ifx{#1\undefined}
}%
\providecommand \@ifnum [1]{%
 \ifnum #1\expandafter \@firstoftwo
 \else \expandafter \@secondoftwo
 \fi
}%
\providecommand \@ifx [1]{%
 \ifx #1\expandafter \@firstoftwo
 \else \expandafter \@secondoftwo
 \fi
}%
\providecommand \natexlab [1]{#1}%
\providecommand \enquote  [1]{``#1''}%
\providecommand \bibnamefont  [1]{#1}%
\providecommand \bibfnamefont [1]{#1}%
\providecommand \citenamefont [1]{#1}%
\providecommand \href@noop [0]{\@secondoftwo}%
\providecommand \href [0]{\begingroup \@sanitize@url \@href}%
\providecommand \@href[1]{\@@startlink{#1}\@@href}%
\providecommand \@@href[1]{\endgroup#1\@@endlink}%
\providecommand \@sanitize@url [0]{\catcode `\\12\catcode `\$12\catcode
  `\&12\catcode `\#12\catcode `\^12\catcode `\_12\catcode `\%12\relax}%
\providecommand \@@startlink[1]{}%
\providecommand \@@endlink[0]{}%
\providecommand \url  [0]{\begingroup\@sanitize@url \@url }%
\providecommand \@url [1]{\endgroup\@href {#1}{\urlprefix }}%
\providecommand \urlprefix  [0]{URL }%
\providecommand \Eprint [0]{\href }%
\providecommand \doibase [0]{https://doi.org/}%
\providecommand \selectlanguage [0]{\@gobble}%
\providecommand \bibinfo  [0]{\@secondoftwo}%
\providecommand \bibfield  [0]{\@secondoftwo}%
\providecommand \translation [1]{[#1]}%
\providecommand \BibitemOpen [0]{}%
\providecommand \bibitemStop [0]{}%
\providecommand \bibitemNoStop [0]{.\EOS\space}%
\providecommand \EOS [0]{\spacefactor3000\relax}%
\providecommand \BibitemShut  [1]{\csname bibitem#1\endcsname}%
\let\auto@bib@innerbib\@empty
\bibitem [{\citenamefont {Strogatz}(2004)}]{strogatz2004sync}%
  \BibitemOpen
  \bibfield  {author} {\bibinfo {author} {\bibfnamefont {S.}~\bibnamefont
  {Strogatz}},\ }\href@noop {} {\emph {\bibinfo {title} {Sync: The emerging
  science of spontaneous order}}}\ (\bibinfo  {publisher} {Penguin UK},\
  \bibinfo {year} {2004})\BibitemShut {NoStop}%
\bibitem [{\citenamefont {Zhang}\ \emph {et~al.}(2020)\citenamefont {Zhang},
  \citenamefont {Cao}, \citenamefont {Ouyang},\ and\ \citenamefont
  {Tu}}]{zhang2020energy}%
  \BibitemOpen
  \bibfield  {author} {\bibinfo {author} {\bibfnamefont {D.}~\bibnamefont
  {Zhang}}, \bibinfo {author} {\bibfnamefont {Y.}~\bibnamefont {Cao}}, \bibinfo
  {author} {\bibfnamefont {Q.}~\bibnamefont {Ouyang}},\ and\ \bibinfo {author}
  {\bibfnamefont {Y.}~\bibnamefont {Tu}},\ }\href@noop {} {\bibfield  {journal}
  {\bibinfo  {journal} {Nature physics}\ }\textbf {\bibinfo {volume} {16}},\
  \bibinfo {pages} {95} (\bibinfo {year} {2020})}\BibitemShut {NoStop}%
\bibitem [{\citenamefont {Xi}\ \emph {et~al.}(2018)\citenamefont {Xi},
  \citenamefont {Wang}, \citenamefont {Liu},\ and\ \citenamefont
  {Wang}}]{xi2018dynamic}%
  \BibitemOpen
  \bibfield  {author} {\bibinfo {author} {\bibfnamefont {J.}~\bibnamefont
  {Xi}}, \bibinfo {author} {\bibfnamefont {C.}~\bibnamefont {Wang}}, \bibinfo
  {author} {\bibfnamefont {H.}~\bibnamefont {Liu}},\ and\ \bibinfo {author}
  {\bibfnamefont {Z.}~\bibnamefont {Wang}},\ }\href@noop {} {\bibfield
  {journal} {\bibinfo  {journal} {IEEE Access}\ }\textbf {\bibinfo {volume}
  {6}},\ \bibinfo {pages} {28923} (\bibinfo {year} {2018})}\BibitemShut
  {NoStop}%
\bibitem [{\citenamefont {Nishikawa}\ and\ \citenamefont
  {Motter}(2006)}]{nishikawa2006maximum}%
  \BibitemOpen
  \bibfield  {author} {\bibinfo {author} {\bibfnamefont {T.}~\bibnamefont
  {Nishikawa}}\ and\ \bibinfo {author} {\bibfnamefont {A.~E.}\ \bibnamefont
  {Motter}},\ }\href@noop {} {\bibfield  {journal} {\bibinfo  {journal}
  {Physica D: Nonlinear Phenomena}\ }\textbf {\bibinfo {volume} {224}},\
  \bibinfo {pages} {77} (\bibinfo {year} {2006})}\BibitemShut {NoStop}%
\bibitem [{\citenamefont {Zhang}\ and\ \citenamefont
  {Strogatz}(2021)}]{zhang2021designing}%
  \BibitemOpen
  \bibfield  {author} {\bibinfo {author} {\bibfnamefont {Y.}~\bibnamefont
  {Zhang}}\ and\ \bibinfo {author} {\bibfnamefont {S.~H.}\ \bibnamefont
  {Strogatz}},\ }\href@noop {} {\bibfield  {journal} {\bibinfo  {journal}
  {Nature communications}\ }\textbf {\bibinfo {volume} {12}},\ \bibinfo {pages}
  {3273} (\bibinfo {year} {2021})}\BibitemShut {NoStop}%
\bibitem [{\citenamefont {Stefanescu}\ \emph {et~al.}(2012)\citenamefont
  {Stefanescu}, \citenamefont {Shivakeshavan},\ and\ \citenamefont
  {Talathi}}]{stefanescu2012computational}%
  \BibitemOpen
  \bibfield  {author} {\bibinfo {author} {\bibfnamefont {R.~A.}\ \bibnamefont
  {Stefanescu}}, \bibinfo {author} {\bibfnamefont {R.}~\bibnamefont
  {Shivakeshavan}},\ and\ \bibinfo {author} {\bibfnamefont {S.~S.}\
  \bibnamefont {Talathi}},\ }\href@noop {} {\bibfield  {journal} {\bibinfo
  {journal} {Seizure}\ }\textbf {\bibinfo {volume} {21}},\ \bibinfo {pages}
  {748} (\bibinfo {year} {2012})}\BibitemShut {NoStop}%
\bibitem [{\citenamefont {Musizza}\ \emph {et~al.}(2007)\citenamefont
  {Musizza}, \citenamefont {Stefanovska}, \citenamefont {McClintock},
  \citenamefont {Palu{\v{s}}}, \citenamefont {Petrov{\v{c}}i{\v{c}}},
  \citenamefont {Ribari{\v{c}}},\ and\ \citenamefont
  {Bajrovi{\'c}}}]{musizza2007interactions}%
  \BibitemOpen
  \bibfield  {author} {\bibinfo {author} {\bibfnamefont {B.}~\bibnamefont
  {Musizza}}, \bibinfo {author} {\bibfnamefont {A.}~\bibnamefont
  {Stefanovska}}, \bibinfo {author} {\bibfnamefont {P.~V.}\ \bibnamefont
  {McClintock}}, \bibinfo {author} {\bibfnamefont {M.}~\bibnamefont
  {Palu{\v{s}}}}, \bibinfo {author} {\bibfnamefont {J.}~\bibnamefont
  {Petrov{\v{c}}i{\v{c}}}}, \bibinfo {author} {\bibfnamefont {S.}~\bibnamefont
  {Ribari{\v{c}}}},\ and\ \bibinfo {author} {\bibfnamefont {F.~F.}\
  \bibnamefont {Bajrovi{\'c}}},\ }\href@noop {} {\bibfield  {journal} {\bibinfo
   {journal} {The journal of Physiology}\ }\textbf {\bibinfo {volume} {580}},\
  \bibinfo {pages} {315} (\bibinfo {year} {2007})}\BibitemShut {NoStop}%
\bibitem [{\citenamefont {Shiogai}\ \emph {et~al.}(2010)\citenamefont
  {Shiogai}, \citenamefont {Stefanovska},\ and\ \citenamefont
  {McClintock}}]{shiogai2010nonlinear}%
  \BibitemOpen
  \bibfield  {author} {\bibinfo {author} {\bibfnamefont {Y.}~\bibnamefont
  {Shiogai}}, \bibinfo {author} {\bibfnamefont {A.}~\bibnamefont
  {Stefanovska}},\ and\ \bibinfo {author} {\bibfnamefont {P.~V.~E.}\
  \bibnamefont {McClintock}},\ }\href@noop {} {\bibfield  {journal} {\bibinfo
  {journal} {Physics reports}\ }\textbf {\bibinfo {volume} {488}},\ \bibinfo
  {pages} {51} (\bibinfo {year} {2010})}\BibitemShut {NoStop}%
\bibitem [{\citenamefont {Suprunenko}\ \emph {et~al.}(2013)\citenamefont
  {Suprunenko}, \citenamefont {Clemson},\ and\ \citenamefont
  {Stefanovska}}]{suprunenko2013chronotaxic}%
  \BibitemOpen
  \bibfield  {author} {\bibinfo {author} {\bibfnamefont {Y.~F.}\ \bibnamefont
  {Suprunenko}}, \bibinfo {author} {\bibfnamefont {P.~T.}\ \bibnamefont
  {Clemson}},\ and\ \bibinfo {author} {\bibfnamefont {A.}~\bibnamefont
  {Stefanovska}},\ }\href@noop {} {\bibfield  {journal} {\bibinfo  {journal}
  {Physical review letters}\ }\textbf {\bibinfo {volume} {111}},\ \bibinfo
  {pages} {024101} (\bibinfo {year} {2013})}\BibitemShut {NoStop}%
\bibitem [{\citenamefont {Lucas}\ \emph {et~al.}(2021)\citenamefont {Lucas},
  \citenamefont {Newman},\ and\ \citenamefont
  {Stefanovska}}]{lucas2021synchronisation}%
  \BibitemOpen
  \bibfield  {author} {\bibinfo {author} {\bibfnamefont {M.}~\bibnamefont
  {Lucas}}, \bibinfo {author} {\bibfnamefont {J.~M.}\ \bibnamefont {Newman}},\
  and\ \bibinfo {author} {\bibfnamefont {A.}~\bibnamefont {Stefanovska}},\ }in\
  \href@noop {} {\emph {\bibinfo {booktitle} {Physics of Biological
  Oscillators}}}\ (\bibinfo  {publisher} {Springer},\ \bibinfo {year} {2021})\
  pp.\ \bibinfo {pages} {85--110}\BibitemShut {NoStop}%
\bibitem [{\citenamefont {Petkoski}\ and\ \citenamefont
  {Stefanovska}(2012)}]{petkoski2012kuramoto}%
  \BibitemOpen
  \bibfield  {author} {\bibinfo {author} {\bibfnamefont {S.}~\bibnamefont
  {Petkoski}}\ and\ \bibinfo {author} {\bibfnamefont {A.}~\bibnamefont
  {Stefanovska}},\ }\href@noop {} {\bibfield  {journal} {\bibinfo  {journal}
  {Physical Review E}\ }\textbf {\bibinfo {volume} {86}},\ \bibinfo {pages}
  {046212} (\bibinfo {year} {2012})}\BibitemShut {NoStop}%
\bibitem [{\citenamefont {Pikovsky}\ \emph {et~al.}(2001)\citenamefont
  {Pikovsky}, \citenamefont {Rosemblum},\ and\ \citenamefont
  {Kurths}}]{pikovsky2001synchronisation}%
  \BibitemOpen
  \bibfield  {author} {\bibinfo {author} {\bibfnamefont {A.}~\bibnamefont
  {Pikovsky}}, \bibinfo {author} {\bibfnamefont {M.}~\bibnamefont
  {Rosemblum}},\ and\ \bibinfo {author} {\bibfnamefont {J.}~\bibnamefont
  {Kurths}},\ }\href@noop {} {\bibfield  {journal} {\bibinfo  {journal} {A
  universal concept in nonlinear sciences. Cambridge Nonlinear Science Series}\
  }\textbf {\bibinfo {volume} {12}} (\bibinfo {year} {2001})}\BibitemShut
  {NoStop}%
\bibitem [{\citenamefont {Gupta}\ \emph {et~al.}(2018)\citenamefont {Gupta},
  \citenamefont {Campa},\ and\ \citenamefont {Ruffo}}]{gupta2018statistical}%
  \BibitemOpen
  \bibfield  {author} {\bibinfo {author} {\bibfnamefont {S.}~\bibnamefont
  {Gupta}}, \bibinfo {author} {\bibfnamefont {A.}~\bibnamefont {Campa}},\ and\
  \bibinfo {author} {\bibfnamefont {S.}~\bibnamefont {Ruffo}},\ }\href@noop {}
  {\emph {\bibinfo {title} {Statistical physics of synchronization}}},\
  Vol.~\bibinfo {volume} {48}\ (\bibinfo  {publisher} {Springer},\ \bibinfo
  {year} {2018})\BibitemShut {NoStop}%
\bibitem [{\citenamefont {Kuramoto}(1984)}]{kuramoto1984chemical}%
  \BibitemOpen
  \bibfield  {author} {\bibinfo {author} {\bibfnamefont {Y.}~\bibnamefont
  {Kuramoto}},\ }in\ \href@noop {} {\emph {\bibinfo {booktitle} {Chemical
  oscillations, waves, and turbulence}}}\ (\bibinfo  {publisher} {Springer},\
  \bibinfo {year} {1984})\ pp.\ \bibinfo {pages} {111--140}\BibitemShut
  {NoStop}%
\bibitem [{\citenamefont {Strogatz}(2000)}]{strogatz2000kuramoto}%
  \BibitemOpen
  \bibfield  {author} {\bibinfo {author} {\bibfnamefont {S.~H.}\ \bibnamefont
  {Strogatz}},\ }\href@noop {} {\bibfield  {journal} {\bibinfo  {journal}
  {Physica D: Nonlinear Phenomena}\ }\textbf {\bibinfo {volume} {143}},\
  \bibinfo {pages} {1} (\bibinfo {year} {2000})}\BibitemShut {NoStop}%
\bibitem [{\citenamefont {Acebr{\'o}n}\ \emph {et~al.}(2005)\citenamefont
  {Acebr{\'o}n}, \citenamefont {Bonilla}, \citenamefont {Vicente},
  \citenamefont {Ritort},\ and\ \citenamefont {Spigler}}]{acebron2005kuramoto}%
  \BibitemOpen
  \bibfield  {author} {\bibinfo {author} {\bibfnamefont {J.~A.}\ \bibnamefont
  {Acebr{\'o}n}}, \bibinfo {author} {\bibfnamefont {L.~L.}\ \bibnamefont
  {Bonilla}}, \bibinfo {author} {\bibfnamefont {C.~J.~P.}\ \bibnamefont
  {Vicente}}, \bibinfo {author} {\bibfnamefont {F.}~\bibnamefont {Ritort}},\
  and\ \bibinfo {author} {\bibfnamefont {R.}~\bibnamefont {Spigler}},\
  }\href@noop {} {\bibfield  {journal} {\bibinfo  {journal} {Reviews of modern
  physics}\ }\textbf {\bibinfo {volume} {77}},\ \bibinfo {pages} {137}
  (\bibinfo {year} {2005})}\BibitemShut {NoStop}%
\bibitem [{\citenamefont {Gupta}\ \emph {et~al.}(2014)\citenamefont {Gupta},
  \citenamefont {Campa},\ and\ \citenamefont {Ruffo}}]{gupta2014kuramoto}%
  \BibitemOpen
  \bibfield  {author} {\bibinfo {author} {\bibfnamefont {S.}~\bibnamefont
  {Gupta}}, \bibinfo {author} {\bibfnamefont {A.}~\bibnamefont {Campa}},\ and\
  \bibinfo {author} {\bibfnamefont {S.}~\bibnamefont {Ruffo}},\ }\href@noop {}
  {\bibfield  {journal} {\bibinfo  {journal} {Journal of Statistical Mechanics:
  Theory and Experiment}\ }\textbf {\bibinfo {volume} {2014}},\ \bibinfo
  {pages} {R08001} (\bibinfo {year} {2014})}\BibitemShut {NoStop}%
\bibitem [{\citenamefont {Verma}\ \emph {et~al.}(2013)\citenamefont {Verma},
  \citenamefont {Contractor},\ and\ \citenamefont
  {Parmananda}}]{verma2013potential}%
  \BibitemOpen
  \bibfield  {author} {\bibinfo {author} {\bibfnamefont {D.~K.}\ \bibnamefont
  {Verma}}, \bibinfo {author} {\bibfnamefont {A.}~\bibnamefont {Contractor}},\
  and\ \bibinfo {author} {\bibfnamefont {P.}~\bibnamefont {Parmananda}},\
  }\href@noop {} {\bibfield  {journal} {\bibinfo  {journal} {The Journal of
  Physical Chemistry A}\ }\textbf {\bibinfo {volume} {117}},\ \bibinfo {pages}
  {267} (\bibinfo {year} {2013})}\BibitemShut {NoStop}%
\bibitem [{\citenamefont {Verma}\ \emph {et~al.}(2015)\citenamefont {Verma},
  \citenamefont {Singh}, \citenamefont {Parmananda}, \citenamefont
  {Contractor},\ and\ \citenamefont {Rivera}}]{verma2015kuramoto}%
  \BibitemOpen
  \bibfield  {author} {\bibinfo {author} {\bibfnamefont {D.~K.}\ \bibnamefont
  {Verma}}, \bibinfo {author} {\bibfnamefont {H.}~\bibnamefont {Singh}},
  \bibinfo {author} {\bibfnamefont {P.}~\bibnamefont {Parmananda}}, \bibinfo
  {author} {\bibfnamefont {A.}~\bibnamefont {Contractor}},\ and\ \bibinfo
  {author} {\bibfnamefont {M.}~\bibnamefont {Rivera}},\ }\href@noop {}
  {\bibfield  {journal} {\bibinfo  {journal} {Chaos: An Interdisciplinary
  Journal of Nonlinear Science}\ }\textbf {\bibinfo {volume} {25}},\ \bibinfo
  {pages} {064609} (\bibinfo {year} {2015})}\BibitemShut {NoStop}%
\bibitem [{\citenamefont {Bera}\ \emph {et~al.}(2017)\citenamefont {Bera},
  \citenamefont {Ghosh}, \citenamefont {Parmananda}, \citenamefont {Osipov},\
  and\ \citenamefont {Dana}}]{bera2017coexisting}%
  \BibitemOpen
  \bibfield  {author} {\bibinfo {author} {\bibfnamefont {B.~K.}\ \bibnamefont
  {Bera}}, \bibinfo {author} {\bibfnamefont {D.}~\bibnamefont {Ghosh}},
  \bibinfo {author} {\bibfnamefont {P.}~\bibnamefont {Parmananda}}, \bibinfo
  {author} {\bibfnamefont {G.}~\bibnamefont {Osipov}},\ and\ \bibinfo {author}
  {\bibfnamefont {S.~K.}\ \bibnamefont {Dana}},\ }\href@noop {} {\bibfield
  {journal} {\bibinfo  {journal} {Chaos: An Interdisciplinary Journal of
  Nonlinear Science}\ }\textbf {\bibinfo {volume} {27}},\ \bibinfo {pages}
  {073108} (\bibinfo {year} {2017})}\BibitemShut {NoStop}%
\bibitem [{\citenamefont {Temirbayev}\ \emph {et~al.}(2013)\citenamefont
  {Temirbayev}, \citenamefont {Nalibayev}, \citenamefont {Zhanabaev},
  \citenamefont {Ponomarenko},\ and\ \citenamefont
  {Rosenblum}}]{temirbayev2013autonomous}%
  \BibitemOpen
  \bibfield  {author} {\bibinfo {author} {\bibfnamefont {A.~A.}\ \bibnamefont
  {Temirbayev}}, \bibinfo {author} {\bibfnamefont {Y.~D.}\ \bibnamefont
  {Nalibayev}}, \bibinfo {author} {\bibfnamefont {Z.~Z.}\ \bibnamefont
  {Zhanabaev}}, \bibinfo {author} {\bibfnamefont {V.~I.}\ \bibnamefont
  {Ponomarenko}},\ and\ \bibinfo {author} {\bibfnamefont {M.}~\bibnamefont
  {Rosenblum}},\ }\href@noop {} {\bibfield  {journal} {\bibinfo  {journal}
  {Physical Review E}\ }\textbf {\bibinfo {volume} {87}},\ \bibinfo {pages}
  {062917} (\bibinfo {year} {2013})}\BibitemShut {NoStop}%
\bibitem [{\citenamefont {English}\ \emph {et~al.}(2015)\citenamefont
  {English}, \citenamefont {Zeng},\ and\ \citenamefont
  {Mertens}}]{english2015experimental}%
  \BibitemOpen
  \bibfield  {author} {\bibinfo {author} {\bibfnamefont {L.~Q.}\ \bibnamefont
  {English}}, \bibinfo {author} {\bibfnamefont {Z.}~\bibnamefont {Zeng}},\ and\
  \bibinfo {author} {\bibfnamefont {D.}~\bibnamefont {Mertens}},\ }\href@noop
  {} {\bibfield  {journal} {\bibinfo  {journal} {Physical Review E}\ }\textbf
  {\bibinfo {volume} {92}},\ \bibinfo {pages} {052912} (\bibinfo {year}
  {2015})}\BibitemShut {NoStop}%
\bibitem [{\citenamefont {Parmananda}\ \emph {et~al.}(2001)\citenamefont
  {Parmananda}, \citenamefont {Mahara}, \citenamefont {Amemiya},\ and\
  \citenamefont {Yamaguchi}}]{parmananda2001resonance}%
  \BibitemOpen
  \bibfield  {author} {\bibinfo {author} {\bibfnamefont {P.}~\bibnamefont
  {Parmananda}}, \bibinfo {author} {\bibfnamefont {H.}~\bibnamefont {Mahara}},
  \bibinfo {author} {\bibfnamefont {T.}~\bibnamefont {Amemiya}},\ and\ \bibinfo
  {author} {\bibfnamefont {T.}~\bibnamefont {Yamaguchi}},\ }\href@noop {}
  {\bibfield  {journal} {\bibinfo  {journal} {Physical review letters}\
  }\textbf {\bibinfo {volume} {87}},\ \bibinfo {pages} {238302} (\bibinfo
  {year} {2001})}\BibitemShut {NoStop}%
\bibitem [{\citenamefont {Chigwada}\ \emph {et~al.}(2006)\citenamefont
  {Chigwada}, \citenamefont {Parmananda},\ and\ \citenamefont
  {Showalter}}]{chigwada2006resonance}%
  \BibitemOpen
  \bibfield  {author} {\bibinfo {author} {\bibfnamefont {T.~R.}\ \bibnamefont
  {Chigwada}}, \bibinfo {author} {\bibfnamefont {P.}~\bibnamefont
  {Parmananda}},\ and\ \bibinfo {author} {\bibfnamefont {K.}~\bibnamefont
  {Showalter}},\ }\href@noop {} {\bibfield  {journal} {\bibinfo  {journal}
  {Physical review letters}\ }\textbf {\bibinfo {volume} {96}},\ \bibinfo
  {pages} {244101} (\bibinfo {year} {2006})}\BibitemShut {NoStop}%
\bibitem [{\citenamefont {Zhang}\ \emph {et~al.}(2021)\citenamefont {Zhang},
  \citenamefont {Ocampo-Espindola}, \citenamefont {Kiss},\ and\ \citenamefont
  {Motter}}]{zhang2021random}%
  \BibitemOpen
  \bibfield  {author} {\bibinfo {author} {\bibfnamefont {Y.}~\bibnamefont
  {Zhang}}, \bibinfo {author} {\bibfnamefont {J.~L.}\ \bibnamefont
  {Ocampo-Espindola}}, \bibinfo {author} {\bibfnamefont {I.~Z.}\ \bibnamefont
  {Kiss}},\ and\ \bibinfo {author} {\bibfnamefont {A.~E.}\ \bibnamefont
  {Motter}},\ }\href@noop {} {\bibfield  {journal} {\bibinfo  {journal}
  {Proceedings of the National Academy of Sciences}\ }\textbf {\bibinfo
  {volume} {118}},\ \bibinfo {pages} {e2024299118} (\bibinfo {year}
  {2021})}\BibitemShut {NoStop}%
\bibitem [{\citenamefont {Sugitani}\ \emph {et~al.}(2021)\citenamefont
  {Sugitani}, \citenamefont {Zhang},\ and\ \citenamefont
  {Motter}}]{sugitani2021synchronizing}%
  \BibitemOpen
  \bibfield  {author} {\bibinfo {author} {\bibfnamefont {Y.}~\bibnamefont
  {Sugitani}}, \bibinfo {author} {\bibfnamefont {Y.}~\bibnamefont {Zhang}},\
  and\ \bibinfo {author} {\bibfnamefont {A.~E.}\ \bibnamefont {Motter}},\
  }\href@noop {} {\bibfield  {journal} {\bibinfo  {journal} {Physical review
  letters}\ }\textbf {\bibinfo {volume} {126}},\ \bibinfo {pages} {164101}
  (\bibinfo {year} {2021})}\BibitemShut {NoStop}%
\bibitem [{\citenamefont {Nicolaou}\ \emph {et~al.}(2020)\citenamefont
  {Nicolaou}, \citenamefont {Sebek}, \citenamefont {Kiss},\ and\ \citenamefont
  {Motter}}]{nicolaou2020coherent}%
  \BibitemOpen
  \bibfield  {author} {\bibinfo {author} {\bibfnamefont {Z.~G.}\ \bibnamefont
  {Nicolaou}}, \bibinfo {author} {\bibfnamefont {M.}~\bibnamefont {Sebek}},
  \bibinfo {author} {\bibfnamefont {I.~Z.}\ \bibnamefont {Kiss}},\ and\
  \bibinfo {author} {\bibfnamefont {A.~E.}\ \bibnamefont {Motter}},\
  }\href@noop {} {\bibfield  {journal} {\bibinfo  {journal} {Physical review
  letters}\ }\textbf {\bibinfo {volume} {125}},\ \bibinfo {pages} {094101}
  (\bibinfo {year} {2020})}\BibitemShut {NoStop}%
\bibitem [{\citenamefont {Nair}\ \emph {et~al.}(2021)\citenamefont {Nair},
  \citenamefont {Hu}, \citenamefont {Berrill}, \citenamefont {Wiesenfeld},\
  and\ \citenamefont {Braiman}}]{nair2021using}%
  \BibitemOpen
  \bibfield  {author} {\bibinfo {author} {\bibfnamefont {N.}~\bibnamefont
  {Nair}}, \bibinfo {author} {\bibfnamefont {K.}~\bibnamefont {Hu}}, \bibinfo
  {author} {\bibfnamefont {M.}~\bibnamefont {Berrill}}, \bibinfo {author}
  {\bibfnamefont {K.}~\bibnamefont {Wiesenfeld}},\ and\ \bibinfo {author}
  {\bibfnamefont {Y.}~\bibnamefont {Braiman}},\ }\href@noop {} {\bibfield
  {journal} {\bibinfo  {journal} {Physical Review Letters}\ }\textbf {\bibinfo
  {volume} {127}},\ \bibinfo {pages} {173901} (\bibinfo {year}
  {2021})}\BibitemShut {NoStop}%
\bibitem [{\citenamefont {Teramae}\ and\ \citenamefont
  {Tanaka}(2004)}]{teramae2004robustness}%
  \BibitemOpen
  \bibfield  {author} {\bibinfo {author} {\bibfnamefont {J.-n.}\ \bibnamefont
  {Teramae}}\ and\ \bibinfo {author} {\bibfnamefont {D.}~\bibnamefont
  {Tanaka}},\ }\href@noop {} {\bibfield  {journal} {\bibinfo  {journal}
  {Physical review letters}\ }\textbf {\bibinfo {volume} {93}},\ \bibinfo
  {pages} {204103} (\bibinfo {year} {2004})}\BibitemShut {NoStop}%
\bibitem [{\citenamefont {Goldobin}\ and\ \citenamefont
  {Pikovsky}(2005)}]{goldobin2005synchronization}%
  \BibitemOpen
  \bibfield  {author} {\bibinfo {author} {\bibfnamefont {D.~S.}\ \bibnamefont
  {Goldobin}}\ and\ \bibinfo {author} {\bibfnamefont {A.}~\bibnamefont
  {Pikovsky}},\ }\href@noop {} {\bibfield  {journal} {\bibinfo  {journal}
  {Physical Review E}\ }\textbf {\bibinfo {volume} {71}},\ \bibinfo {pages}
  {045201} (\bibinfo {year} {2005})}\BibitemShut {NoStop}%
\bibitem [{\citenamefont {Nakao}\ \emph {et~al.}(2007)\citenamefont {Nakao},
  \citenamefont {Arai},\ and\ \citenamefont {Kawamura}}]{nakao2007noise}%
  \BibitemOpen
  \bibfield  {author} {\bibinfo {author} {\bibfnamefont {H.}~\bibnamefont
  {Nakao}}, \bibinfo {author} {\bibfnamefont {K.}~\bibnamefont {Arai}},\ and\
  \bibinfo {author} {\bibfnamefont {Y.}~\bibnamefont {Kawamura}},\ }\href@noop
  {} {\bibfield  {journal} {\bibinfo  {journal} {Physical review letters}\
  }\textbf {\bibinfo {volume} {98}},\ \bibinfo {pages} {184101} (\bibinfo
  {year} {2007})}\BibitemShut {NoStop}%
\bibitem [{\citenamefont {Yoshimura}\ \emph {et~al.}(2007)\citenamefont
  {Yoshimura}, \citenamefont {Valiusaityte},\ and\ \citenamefont
  {Davis}}]{yoshimura2007synchronization}%
  \BibitemOpen
  \bibfield  {author} {\bibinfo {author} {\bibfnamefont {K.}~\bibnamefont
  {Yoshimura}}, \bibinfo {author} {\bibfnamefont {I.}~\bibnamefont
  {Valiusaityte}},\ and\ \bibinfo {author} {\bibfnamefont {P.}~\bibnamefont
  {Davis}},\ }\href@noop {} {\bibfield  {journal} {\bibinfo  {journal}
  {Physical Review E}\ }\textbf {\bibinfo {volume} {75}},\ \bibinfo {pages}
  {026208} (\bibinfo {year} {2007})}\BibitemShut {NoStop}%
\bibitem [{\citenamefont {Nagai}\ and\ \citenamefont
  {Kori}(2010)}]{nagai2010noise}%
  \BibitemOpen
  \bibfield  {author} {\bibinfo {author} {\bibfnamefont {K.~H.}\ \bibnamefont
  {Nagai}}\ and\ \bibinfo {author} {\bibfnamefont {H.}~\bibnamefont {Kori}},\
  }\href@noop {} {\bibfield  {journal} {\bibinfo  {journal} {Physical Review
  E}\ }\textbf {\bibinfo {volume} {81}},\ \bibinfo {pages} {065202} (\bibinfo
  {year} {2010})}\BibitemShut {NoStop}%
\bibitem [{\citenamefont {Pinto}\ \emph {et~al.}(2016)\citenamefont {Pinto},
  \citenamefont {Oliveira},\ and\ \citenamefont
  {Penna}}]{pinto2016thermodynamics}%
  \BibitemOpen
  \bibfield  {author} {\bibinfo {author} {\bibfnamefont {P.~D.}\ \bibnamefont
  {Pinto}}, \bibinfo {author} {\bibfnamefont {F.~A.}\ \bibnamefont
  {Oliveira}},\ and\ \bibinfo {author} {\bibfnamefont {A.~L.}\ \bibnamefont
  {Penna}},\ }\href@noop {} {\bibfield  {journal} {\bibinfo  {journal}
  {Physical Review E}\ }\textbf {\bibinfo {volume} {93}},\ \bibinfo {pages}
  {052220} (\bibinfo {year} {2016})}\BibitemShut {NoStop}%
\bibitem [{\citenamefont {Kawamura}\ and\ \citenamefont
  {Nakao}(2016)}]{kawamura2016optimization}%
  \BibitemOpen
  \bibfield  {author} {\bibinfo {author} {\bibfnamefont {Y.}~\bibnamefont
  {Kawamura}}\ and\ \bibinfo {author} {\bibfnamefont {H.}~\bibnamefont
  {Nakao}},\ }\href@noop {} {\bibfield  {journal} {\bibinfo  {journal}
  {Physical Review E}\ }\textbf {\bibinfo {volume} {94}},\ \bibinfo {pages}
  {032201} (\bibinfo {year} {2016})}\BibitemShut {NoStop}%
\bibitem [{\citenamefont {Cox}(1962)}]{cox1962renewal}%
  \BibitemOpen
  \bibfield  {author} {\bibinfo {author} {\bibfnamefont {D.~R.}\ \bibnamefont
  {Cox}},\ }\href@noop {} {\bibfield  {journal} {\bibinfo  {journal} {Methuen
  and Co. Ltd., London}\ } (\bibinfo {year} {1962})}\BibitemShut {NoStop}%
\bibitem [{not()}]{note}%
  \BibitemOpen
  \href@noop {} {}\bibinfo {note} {We have checked in simulations (data not
  presented here) that qualitatively similar results follow for the
  representative class of power-law $g(\omega)$, namely, distributions with
  power-law tails: $g(\omega) \sim \omega^{-n}$ for large $|\omega|$, so long
  as the variance is finite, i.e., for $n>3$.}\BibitemShut {Stop}%
\bibitem [{\citenamefont {Fisher}(1967)}]{Fisher1967}%
  \BibitemOpen
  \bibfield  {author} {\bibinfo {author} {\bibfnamefont {M.~E.}\ \bibnamefont
  {Fisher}},\ }\href@noop {} {\bibfield  {journal} {\bibinfo  {journal}
  {Reports on Progress in Physics}\ }\textbf {\bibinfo {volume} {30}},\
  \bibinfo {pages} {615} (\bibinfo {year} {1967})}\BibitemShut {NoStop}%
\bibitem [{\citenamefont {Landau}\ and\ \citenamefont
  {Binder}(2009)}]{binder-book}%
  \BibitemOpen
  \bibfield  {author} {\bibinfo {author} {\bibfnamefont {D.~P.}\ \bibnamefont
  {Landau}}\ and\ \bibinfo {author} {\bibfnamefont {K.}~\bibnamefont
  {Binder}},\ }\href@noop {} {\emph {\bibinfo {title} {A Guide to Monte Carlo
  Simulations in Statistical Physics}}}\ (\bibinfo  {publisher} {Cambridge
  University Press},\ \bibinfo {year} {2009})\BibitemShut {NoStop}%
\bibitem [{\citenamefont {English}\ \emph {et~al.}(2016)\citenamefont
  {English}, \citenamefont {Mertens}, \citenamefont {Abdoulkary}, \citenamefont
  {Fritz}, \citenamefont {Skowronski},\ and\ \citenamefont
  {Kevrekidis}}]{english2016emergence}%
  \BibitemOpen
  \bibfield  {author} {\bibinfo {author} {\bibfnamefont {L.~Q.}\ \bibnamefont
  {English}}, \bibinfo {author} {\bibfnamefont {D.}~\bibnamefont {Mertens}},
  \bibinfo {author} {\bibfnamefont {S.}~\bibnamefont {Abdoulkary}}, \bibinfo
  {author} {\bibfnamefont {C.~B.}\ \bibnamefont {Fritz}}, \bibinfo {author}
  {\bibfnamefont {K.}~\bibnamefont {Skowronski}},\ and\ \bibinfo {author}
  {\bibfnamefont {P.}~\bibnamefont {Kevrekidis}},\ }\href@noop {} {\bibfield
  {journal} {\bibinfo  {journal} {Physical Review E}\ }\textbf {\bibinfo
  {volume} {94}},\ \bibinfo {pages} {062212} (\bibinfo {year}
  {2016})}\BibitemShut {NoStop}%
\bibitem [{\citenamefont {Melby}\ \emph {et~al.}(2005)\citenamefont {Melby},
  \citenamefont {Weber},\ and\ \citenamefont {H{\"u}bler}}]{melby2005dynamics}%
  \BibitemOpen
  \bibfield  {author} {\bibinfo {author} {\bibfnamefont {P.}~\bibnamefont
  {Melby}}, \bibinfo {author} {\bibfnamefont {N.}~\bibnamefont {Weber}},\ and\
  \bibinfo {author} {\bibfnamefont {A.}~\bibnamefont {H{\"u}bler}},\
  }\href@noop {} {\bibfield  {journal} {\bibinfo  {journal} {Chaos: An
  Interdisciplinary Journal of Nonlinear Science}\ }\textbf {\bibinfo {volume}
  {15}},\ \bibinfo {pages} {033902} (\bibinfo {year} {2005})}\BibitemShut
  {NoStop}%
\end{thebibliography}

\end{document}